\documentclass[11pt]{article}
\usepackage{graphicx}
\usepackage{amssymb,amsmath,amsfonts}

\def\Re{{\cal R \mskip-4mu \lower.1ex \hbox{\it e}\,}}
\def\Im{{\cal I \mskip-5mu \lower.1ex \hbox{\it m}\,}}
\def\ie{{\it i.e.}}
\def\eg{{\it e.g.}}

\def\sub#1{_{\lower.25ex\hbox{$\scriptstyle#1$}}}
\def\tev{\,{\ifmmode\mathrm {TeV}\else TeV\fi}}
\def\gev{\,{\ifmmode\mathrm {GeV}\else GeV\fi}}
\def\mev{\,{\ifmmode\mathrm {MeV}\else MeV\fi}}
\def\mpl{\ifmmode M_{pl}\else $M_{pl}$\fi}
\def\mpl{\ifmmode \overline M_{Pl}\else $\bar M_{Pl}$\fi}
\def\to{\rightarrow}

\def\subw{_{\rm w}}
\def\mh{\ifmmode m\sbl H \else $m\sbl H$\fi}
\def\mch{\ifmmode m_{H^\pm} \else $m_{H^\pm}$\fi}
\def\mt{\ifmmode m_t\else $m_t$\fi}
\def\mc{\ifmmode m_c\else $m_c$\fi}
\def\mz{\ifmmode M_Z\else $M_Z$\fi}
\def\mw{\ifmmode M_W\else $M_W$\fi}
\def\mws{\ifmmode M_W^2 \else $M_W^2$\fi}
\def\mhs{\ifmmode m_H^2 \else $m_H^2$\fi}   
\def\mzs{\ifmmode M_Z^2 \else $M_Z^2$\fi}
\def\mts{\ifmmode m_t^2 \else $m_t^2$\fi}
\def\mcs{\ifmmode m_c^2 \else $m_c^2$\fi}
\def\mchs{\ifmmode m_{H^\pm}^2 \else $m_{H^\pm}^2$\fi}
\def\ztwo{\ifmmode Z_2\else $Z_2$\fi}
\def\zone{\ifmmode Z_1\else $Z_1$\fi}
\def\mtwo{\ifmmode M_2\else $M_2$\fi}
\def\mone{\ifmmode M_1\else $M_1$\fi}
\def\tb{\ifmmode \tan\beta \else $\tan\beta$\fi}
\def\xw{\ifmmode x\subw\else $x\subw$\fi}
\def\ch{\ifmmode H^\pm \else $H^\pm$\fi}
\def\lum{\ifmmode {\cal L}\else ${\cal L}$\fi}
\def\inpb{\,{\ifmmode {\mathrm {pb}}^{-1}\else ${\mathrm {pb}}^{-1}$\fi}}
\def\infb{\,{\ifmmode {\mathrm {fb}}^{-1}\else ${\mathrm {fb}}^{-1}$\fi}}
\def\epem{\ifmmode e^+e^-\else $e^+e^-$\fi}
\def\ppb{\ifmmode \bar pp\else $\bar pp$\fi}
\def\bsg{\ifmmode B\to X_s\gamma\else $B\to X_s\gamma$\fi}
\def\bsll{\ifmmode B\to X_s\ell^+\ell^-\else $B\to X_s\ell^+\ell^-$\fi}
\def\bstt{\ifmmode B\to X_s\tau^+\tau^-\else $B\to X_s\tau^+\tau^-$\fi}
\def\lamt{\ifmmode \tilde\lambda\else $\tilde\lambda$\fi}
\def\shat{\ifmmode \hat s\else $\hat s$\fi}
\def\that{\ifmmode \hat t\else $\hat t$\fi}
\def\uhat{\ifmmode \hat u\else $\hat u$\fi}

\newskip\zatskip \zatskip=0pt plus0pt minus0pt
\def\matth{\mathsurround=0pt}
\def\lsim{\mathrel{\mathpalette\atversim<}}

\def\atversim#1#2{\lower0.7ex\vbox{\baselineskip\zatskip\lineskip\zatskip
  \lineskiplimit 0pt\ialign{$\matth#1\hfil##\hfil$\crcr#2\crcr\sim\crcr}}}

\def\grtsim{\,\,\rlap{\raise 3pt\hbox{$>$}}{\lower 3pt\hbox{$\sim$}}\,\,}
\def\lsim{\,\,\rlap{\raise 3pt\hbox{$<$}}{\lower 3pt\hbox{$\sim$}}\,\,}

\renewcommand{\thefootnote}{\fnsymbol{footnote}}

\begin{document} \begin{titlepage}
\rightline{\vbox{\halign{&#\hfil\cr
&SLAC-PUB-13039\cr
}}}
\begin{center}
\thispagestyle{empty} \flushbottom { {
\Large\bf Unique Identification of Lee-Wick Gauge Bosons at Linear Colliders 
\footnote{Work supported in part
by the Department of Energy, Contract DE-AC02-76SF00515}
\footnote{e-mail:
rizzo@slac.stanford.edu}}}
\medskip
\end{center}

\centerline{Thomas G. Rizzo}
\vspace{8pt} 
\centerline{\it Stanford Linear
Accelerator Center, 2575 Sand Hill Rd., Menlo Park, CA, 94025}

\vspace*{0.3cm}

\begin{abstract}
Grinstein, O'Connell and Wise have recently presented an extension of the Standard Model (SM), based on the ideas of Lee and Wick (LW), 
which demonstrates an interesting way to remove the quadratically divergent contributions to the Higgs mass induced by radiative corrections. 
This model predicts the existence of negative-norm copies of the usual SM fields at the TeV scale with ghost-like propagators and negative 
decay widths, but with otherwise SM-like couplings. In earlier work, it was demonstrated that the LW states in the gauge boson sector of these 
models, though easy to observe, cannot be uniquely identified as such at the LHC. In this paper, we address the issue of whether or not 
this problem can be resolved at an $e^+e^-$ collider with a suitable center of mass energy range. We find that measurements of the cross section 
and the left-right polarization asymmetry associated with Bhabha scattering can lead to a unique identification of the neutral electroweak 
gauge bosons of the Lee-Wick type.   
\end{abstract}



\renewcommand{\thefootnote}{\arabic{footnote}} \end{titlepage} 

%
%
%

\section{Introduction and Background}

One of the outstanding problems facing high energy physics is the origin of electroweak symmetry breaking. Although the usual Higgs mechanism, which 
employs a single weak scalar isodoublet, is phenomenologically successful{\cite {LEPEWWG}} it is not theoretically satisfying.  Is it possible to generate 
the masses for the gauge bosons and fermions of the Standard Model(SM) without encountering fine-tuning and naturalness issues as well as the associated hierarchy 
problem? On the experimental side, we expect that the LHC should begin to probe for answers to these important questions over the next few years with potentially 
surprising results. While we wait, it is important for us to examine as many scenarios as possible which address these issues in order to prepare ourselves 
for these critically important Terascale experimental results. 

Grinstein, O'Connell and Wise(GOW){\cite {GOW}} have recently extended to the SM context an old idea by Lee and Wick(LW){\cite {LW}} based on higher-derivative 
theories. This model apparently solves the hierarchy problem and eliminates the quadratic divergence of the Higgs boson mass that one 
ordinarily encounters in the SM. The most essential feature of the GOW scenario is the introduction of negative-normed states into the usual SM Hilbert space. In  
particular, one introduces a new massive degree of freedom (or one vector-like pair in the fermion case) for each of the conventional SM particles with the same 
spin. The resulting contributions of these exotic new particles to the Higgs mass quadratic divergence then cancels those of the SM, partner by partner, leaving only 
safe logarithmic terms. In the gauge sector, \eg, the following new fields are introduced: an $SU(3)_c$ octet of `gluons', $g_{LW}$, with mass $M_3$, an $SU(2)_L$ 
isotriplet of weak bosons, $W_{LW}^{0,\pm}$, of mass $M_2$ and a heavy neutral $U(1)_Y$ hypercharge field, $B_{LW}$, with mass $M_1$. The interactions of these new 
fields with each other and with the familiar ones of the SM are given in detail in Ref.{\cite {GOW}}. GOW argue that due to naturalness requirements and the present 
direct{\cite {limit}} and indirect{\cite {LEPEWWG}} experimental constraints on the existence of such particles, one should anticipate that their masses must 
lie not too far above $\simeq 1$ TeV. The implications of such a scenario have been examined in a number of recent works{\cite {big}}.

Within this context, the purpose of the present paper is to address a purely phenomenological issue. As long as such states 
are not too massive, since their interactions are very similar to those of their conventional SM counterparts, it is already clear that they will be produced 
{\it and} observed at the LHC based on the results of other existing analyses{\cite {tdr}}. Due to their rather strange and unusual properties, one might imagine 
that it would be rather trivial for LHC experimental 
data to be used to uniquely identify such states as arising from the GOW framework. However, it was shown in an earlier 
work{\cite {old}} that this is not the case in the gauge boson sector due to the possible ambiguities in the signs of the couplings of new gauge bosons to the 
SM fermions.  The issue we want to address in this paper is whether 
or not this situation can be overcome at future $e^+e^-$ colliders, \ie, can we tell that we have unambiguously observed these negative-metric LW fields and 
not something else? We will demonstrate that measurements of the Bhabha scattering process will allow us to answer this question conclusively in the affirmative.

Since we will be considering $e^+e^-$ collisions, our attention will be focused on the new neutral electroweak gauge bosons in the GOW model. The essential 
phenomenological features of these new states is straightforward to summarize: ($i$) the propagators and decay widths of the relevant LW particles, 
$W_{LW}^0,B_{LW}$, have signs which are opposite to those of the familiar SM fields; ($ii$) the couplings of these LW gauge fields to SM fermions are exactly 
those of the corresponding SM gauge fields; ($iii$) in the limit that $M_{1,2}^2>>M_{W,Z}^2$, as will be the case discussed below, the mixing between the SM 
and GOW gauge bosons can be neglected. When ($i$) and ($ii$) are taken together they imply a strong destructive interference between the SM and GOW amplitudes 
that can be symbolically written as  
\begin{equation}
\sim {{i}\over {p^2-M_{SM}^2+iM_{SM}\Gamma_{SM}}}-{{i}\over {p^2-M_{LW}^2+iM_{LW}\Gamma_{LW}}}\,,
\end{equation}
apart from other overall factors. In particular the width $\Gamma_{LW}<0$ has exactly the same magnitude as would a heavy copy of the relevant SM gauge field. Note 
that here we have assumed that the decays of these heavy gauge bosons into pairs of the LW partners of the SM fermions is not kinematically allowed. In this case, 
the width to mass ratio of these new gauge bosons is quite small $\sim 3\%$. If such decays 
are allowed, only the widths of the new gauge states are modified and not their couplings to the SM fields which is what we wish to explore below. If decays to 
some of these fermions are allowed, we would still expect that $\Gamma_{LW}/M_{LW} \leq 5\%$ or so.  This overall situation is 
somewhat reminiscent of what happens in the Sequential SM(SSM){\cite {tasi}} or the case of flat, TeV-scale extra dimensions where the fermions are confined 
to the origin of the fifth dimension (apart from an additional numerical factor{\cite {flat}} of $\sqrt 2$ which might be modified by the existence of brane-localized 
kinetic terms{\cite {brane}}). A small, but important, difference here is that for the  
general case when $M_1 \neq M_2$, the two fields $W_{LW}^0$ and $B_{LW}$ will be the true mass eigenstates and we shall generally 
use this basis in what follows. To see this, we note that the angle describing the mixing between these two states is given by{\cite {GOW}}
\begin{equation}
\tan 2\phi={{gg'v^2}\over {2}}\Big[M_1^2-M_2^2+(g^2-g'^2){{v^2}\over {2}}\Big]^{-1}\,,
\end{equation}
where $g,g'$ are the usual $SU(2)_L$ and $U(1)_Y$ SM couplings with $v$ the SM Higgs vev. When $M_1$ and $M_2$ are substantially different, this mixing is quite small, 
\ie, of order $10^{-2}$ or less. However, when $M_1=M_2$, a special case that we will consider below, the angle $\phi$ is large and is seen to be identical with the 
usual weak mixing angle, $\theta_w$. Clearly, mixing must be included in this case in any phenomenological analysis.

Before beginning our analysis let us remind the reader about the source of the ambiguities encountered at the LHC. Due to our particular interests below we focus 
on the 
the neutral electroweak gauge boson sector though the same problems arise for all gauge sectors. The primary way to observe a new gauge field with SM-like couplings 
at the LHC is in the Drell-Yan channel{\cite {tdr,tasi}}. As an example, let us consider the production and decay of heavy $W^0,B$-like states at the LHC, 
comparing three possibilities: $(a)$ the new fields are exact but heavier duplicates of the ones in the SM and, as discussed above, might occur in models with 
extra dimensions, $(b)$ they are LW-type fields, or $(c)$ they are SM-like 
fields {\it but} the overall relative sign between, \eg, the initial state quarks and the final state leptonic couplings is opposite to that of the SM. As discussed 
in our earlier work{\cite {old}}, it was noted that such a situation could arise in models{\cite {nima}} where fermions are localized on two different branes bounding 
an extra flat dimension. Note that for the following phenomemological discussion, these alternatives to the LW model are treated only as `strawmen' against which the 
LW predictions can be tested. In the  
resonance region(s) these three scenarios are essentially identical producing resonances with exactly the same (apparent) widths and branching fractions and with the 
same angular distributions for the outgoing leptons. Below the resonances, $(a)$ differs from $(b)$ and $(c)$ since there is strong destructive interference in 
this case whereas the other two scenarios lead to constructive interference with the SM photon and $Z$ exchanges. Thus case $(a)$ can be  
distinguished from cases $(b)$ and $(c)$ by measuring the cross section in this interference region. However, cases $(b)$ and $(c)$ are found to be indistinguishable; 
algebraically, the corresponding amplitudes in these two cases differ only in the sign of the imaginary parts in the $W_{LW}^0$ and $B_{LW}$ contributions which are 
sufficiently small in comparison to other terms in the amplitude as to be impossible to observe{\cite {old}} even at very high 
LHC integrated luminosities. Can we get around this problem at an $e^+e^-$ collider and separate all three of these possibilities, uniquely establishing the identity 
of the LW states? This is the issue to which we now turn.

\section{Analysis}

To begin our analysis and to be as general as possible let us first imagine that we have available to us an $e^+e^-$ collider with an adjustable center of mass energy 
in the TeV range which will follow in the wake of the LHC. 
Consider the set of processes $e^+e^-\to f\bar f$ where $f$ is any SM fermion. Then it is well known{\cite {tasi}} that for any (massless) 
fermion, $f\neq e$, the Born-level production differential 
cross section for unpolarized $e^\pm$ due to the $s$-channel exchange of any number of (ordinary) neutral gauge bosons can be written as 
\begin{equation}
{{d\sigma}\over {dz}}={N_c\over {32\pi s}}\sum_{i,j}P^{ss}_{ij}[B_{ij}(1+z^2)+2C_{ij}z]\,,
\end{equation}
where $N_c$ is a color factor, $z=\cos \theta$, the angle being between $e^-$ and $f$, with  
\begin{eqnarray}
B_{ij}&=&(v_iv_j+a_ia_j)_e(v_iv_j+a_ia_j)_f\\ \nonumber
C_{ij}&=&(v_ia_j+a_iv_j)_e(v_ia_j+a_iv_j)_f\,,
\end{eqnarray}
with $v_i,a_i$ being the vector and axial vector couplings of $e$ and $f$ to the $i$th gauge boson and
\begin{equation}
P^{ss}_{ij}=s^2{{(s-M_i^2)(s-M_j^2)+\Gamma_i\Gamma_jM_iM_j}\over {[(s-M_i^2)^2+\Gamma_i^2M_i^2][i\to j]}}\,,
\end{equation}
is the corresponding propagator factor. Here $M_i(\Gamma_i)$ are the mass (width) of the $i$th gauge boson.  For polarized beams, a similar set of expressions can be 
written down to construct the left-right polarization asymmetry, $A_{LR}^f(z)$; to do this we make the replacements   
\begin{eqnarray}
B_{ij}&\to& B_{ij}+\xi (v_ia_j+a_iv_j)_e(v_iv_j+a_ia_j)_f\\ \nonumber 
C_{ij}&\to& C_{ij}+\xi (v_iv_j+a_ia_j)_e(v_ia_j+a_iv_j)_f\,,
\end{eqnarray}
and then form the ratio 
\begin{equation}
A_{LR}^f(z)=P\Bigg[{{d\sigma(\xi=+1)-d\sigma(\xi=-1)}\over {~~~~''~~~~+~~~~''~~~~}}\Bigg]\,, 
\end{equation}
where $P$ is the effective beam polarization. In the calculations below we will set $P=1$ for simplicity.  

\begin{figure}[htbp]
\centerline{
\includegraphics[width=7.5cm,angle=90]{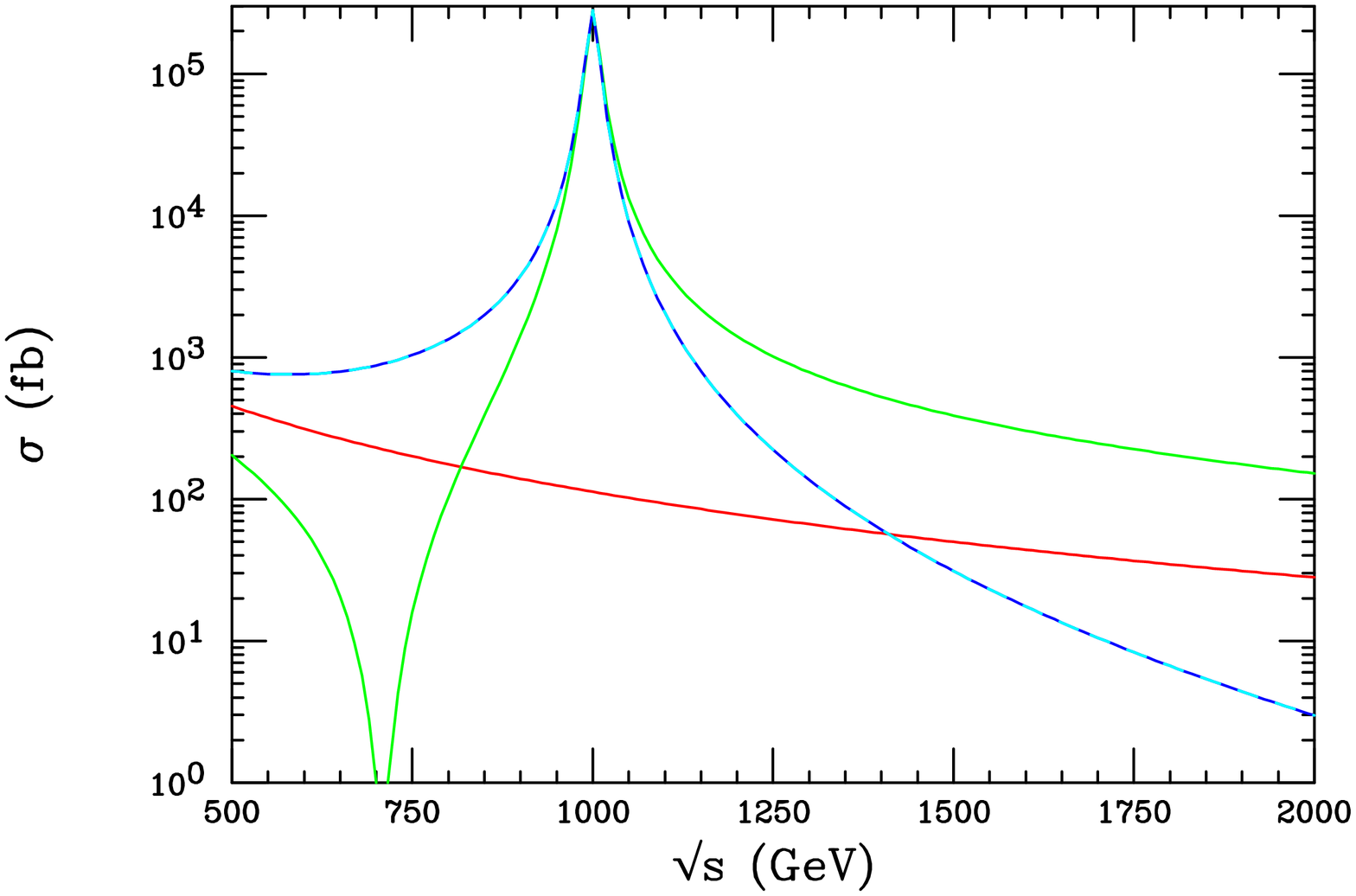}}
\vspace*{0.1cm}
\centerline{
\includegraphics[width=7.5cm,angle=90]{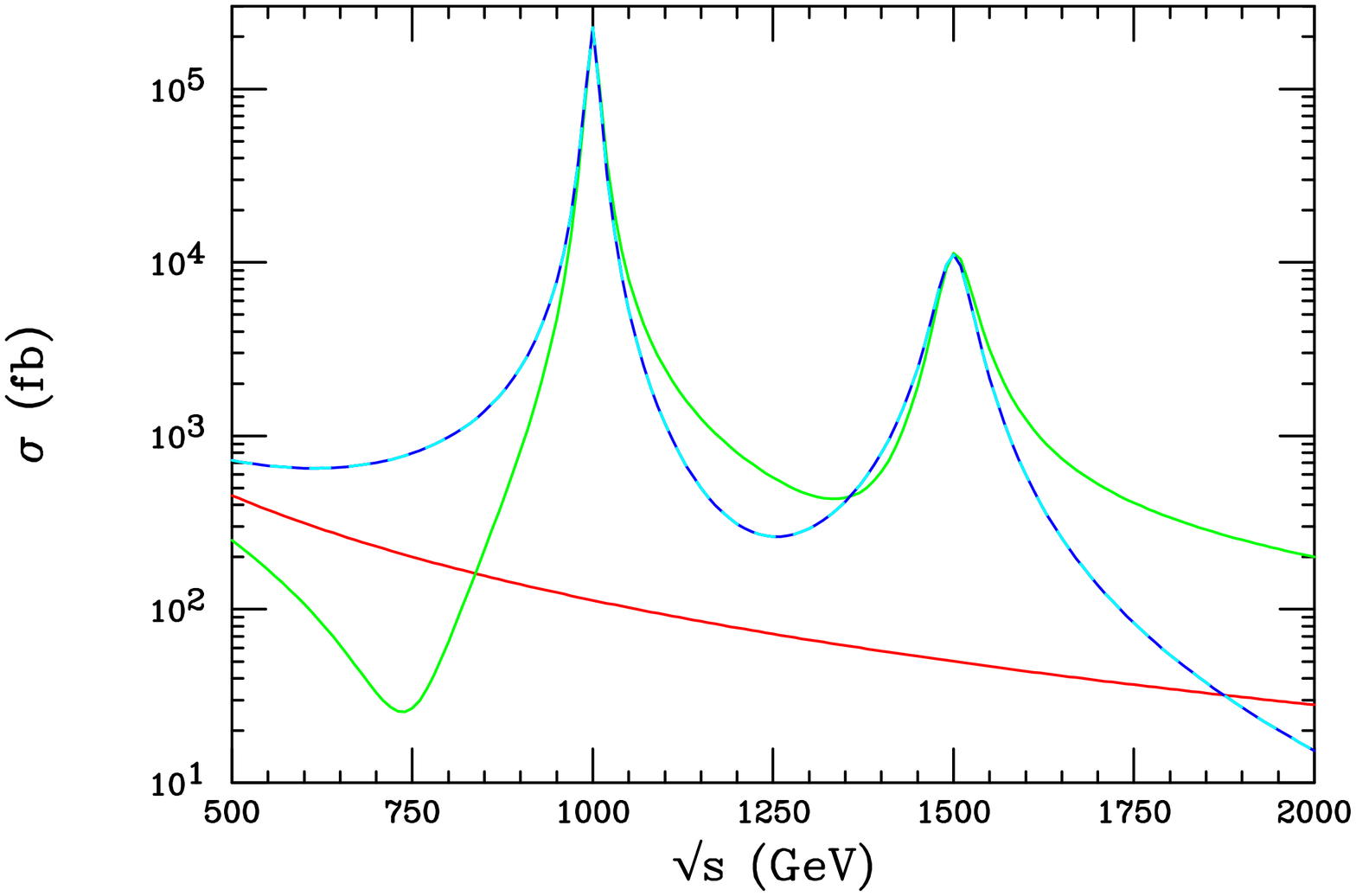}}
\vspace*{0.1cm}
\caption{Cross section for $e^+e^- \to \mu^+\mu^-$ as a function of $\sqrt s$ for scenario $(a)$ in green and for scenarios $(b)$ and $(c)$ in blue. 
The explicit GOW results are shown as dashes inside of the blue curve. The SM prediction 
for comparison purposes is in red. In the top panel $M_{W^0}=M_B=1$ TeV whereas in the bottom panel $M_B=1$ TeV and $M_{W^0}=1.5$ TeV.}
\label{fig1}
\end{figure}

\begin{figure}[htbp]
\centerline{
\includegraphics[width=7.5cm,angle=90]{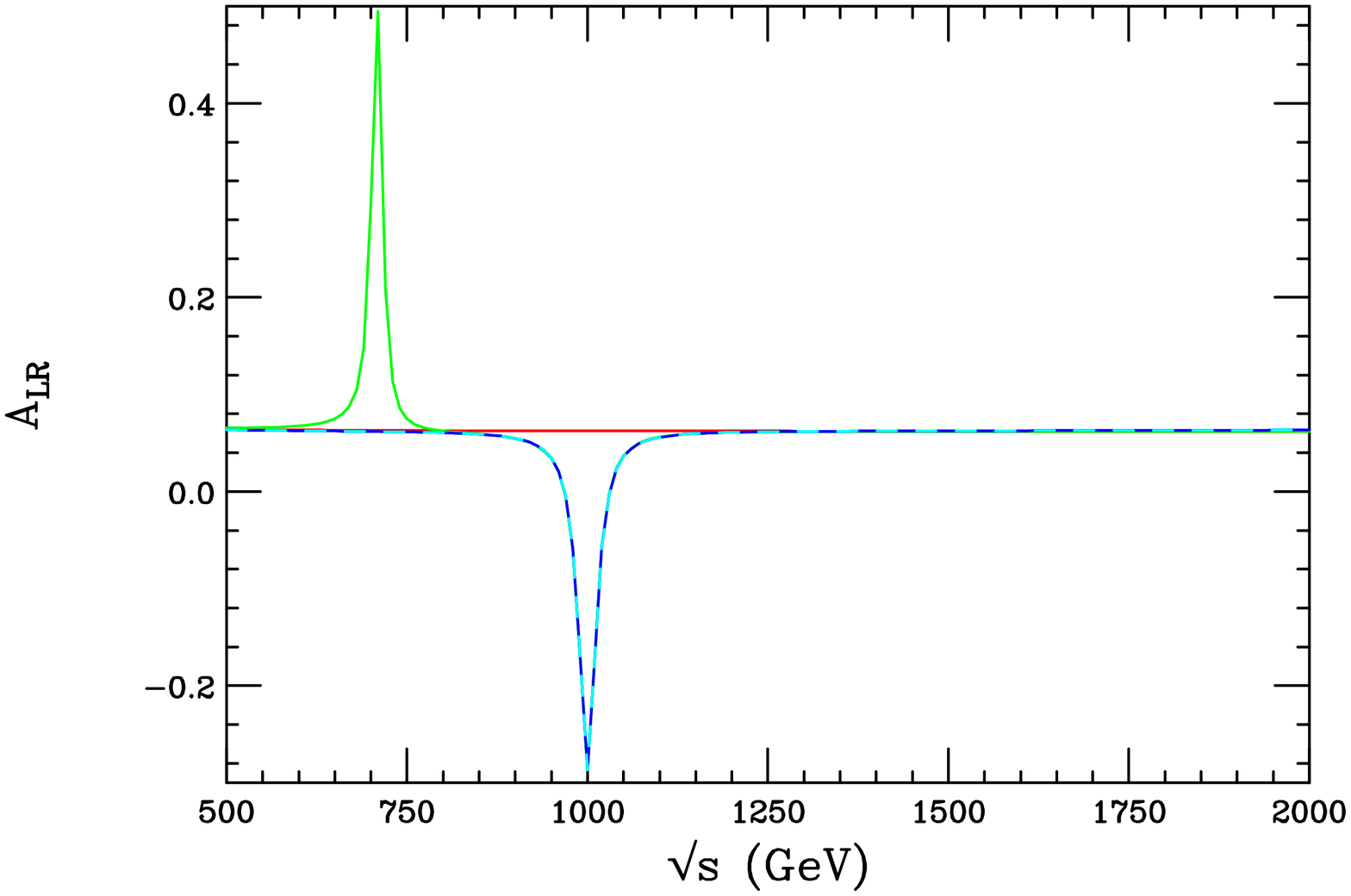}}
\vspace*{0.1cm}
\centerline{
\includegraphics[width=7.5cm,angle=90]{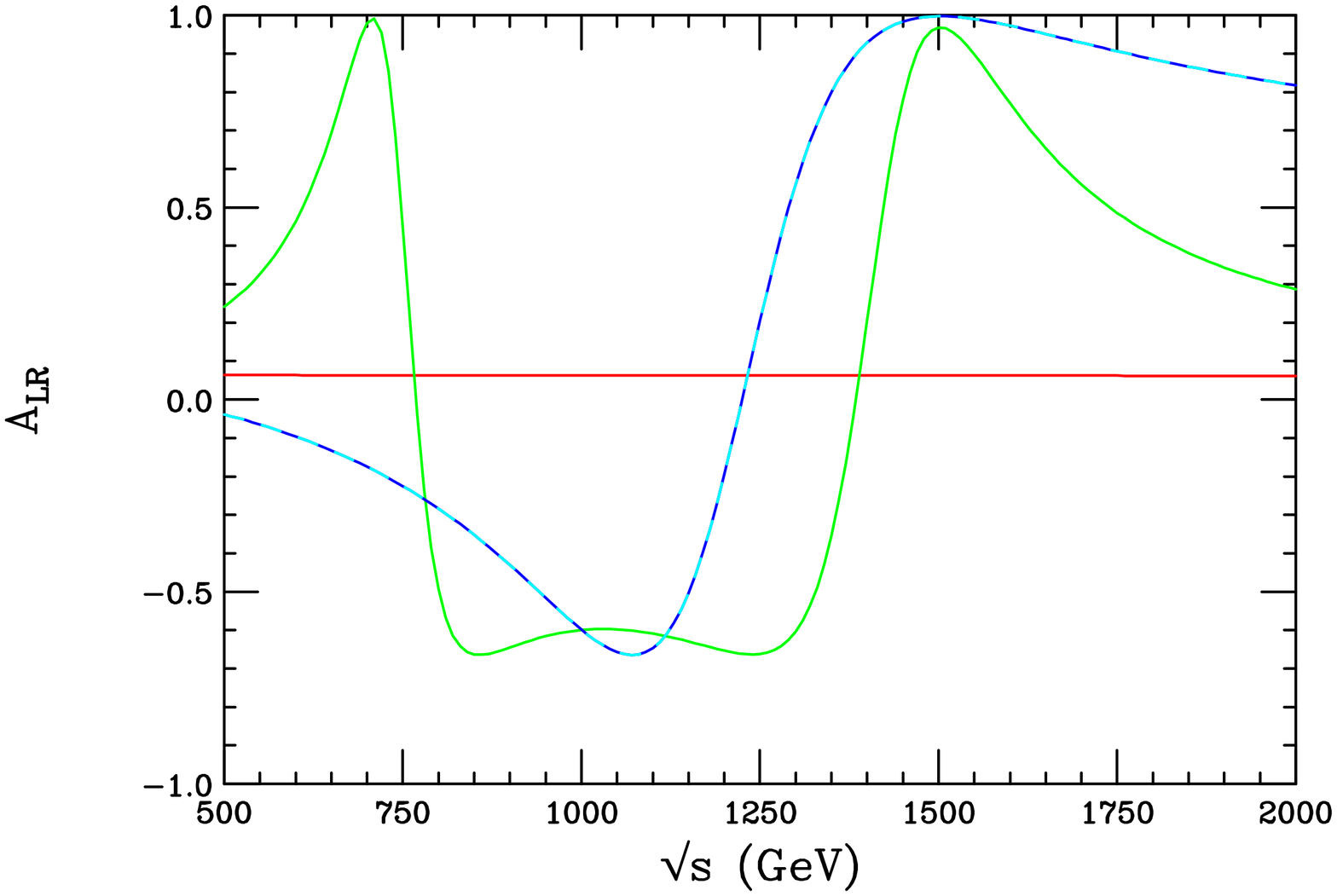}}
\vspace*{0.1cm}
\caption{Same as the previous figure but now for the angular averaged values of $A_{LR}^{\mu}$.}
\label{fig1a}
\end{figure}

Let us now consider the three models $(a)-(c)$ in this environment; the expressions 
above apply directly to cases $(a)$ and $(c)$ as the gauge fields in these two cases are `ordinary'. As at the LHC, we see that flipping the relative sign of the 
initial/final fermion couplings of the $W^0$ and $B$ will change the the cross section in the interference region both below and above the resonances. This is shown 
explicitly in Figs.~\ref{fig1} and ~\ref{fig1a} for two representative spectrum cases assuming for simplicity that $f=\mu$. To cover the case of LW gauge bosons, we 
must recall that now $\Gamma_{W^0,B}<0$ and rescale the equation for the  $P^{ss}_{ij}$: $P^{ss}_{ij}\to \lambda_{ij}P^{ss}_{ij}$, where $\lambda_{ij}=1$ when 
both $i,j$ both correspond to SM or LW gauge fields but $=-1$ in all other cases where SM and LW exchanges interfere. 
It is clear from this exercise that the cross sections for scenarios $(b)$ and $(c)$ will differ by construction only in the sign of 
the terms proportional to the products $\Gamma_Z\Gamma_{W^0,B}$. Note that the resulting cross section for the LW case, $(b)$, is also shown in  Fig.~\ref{fig1} 
lying directly on top of that for scenario $(c)$, repeating our LHC experience. We also find that a similar result is also observed to hold in the case of the  
angular-averaged values of $A_{LR}^f$, \ie, cases $(b)$ and $(c)$ lead to virtually identical numerical results for $A_{LR}^f$. 


It is clear that we can always play this game with the signs of the couplings on the new gauge bosons 
when the initial state and final state fermions are {\it different}. At the LHC, we attempted{\cite {old}} to circumvent this problem by looking at reactions in 
the QCD sector such as $q\bar q \to q\bar q$, which in this scenario is now also mediated by heavy LW gluons, and which produces the dijet final 
state. Here, the initial and final state partons are {\it identical}. The problem in such a case is that there are many processes which mediate dijet production, 
even at leading order. We showed in that work that is was essentially impossible to isolate the effects of the negative-normed LW exchanges.

\begin{figure}[htbp]
\centerline{
\includegraphics[width=7.5cm,angle=90]{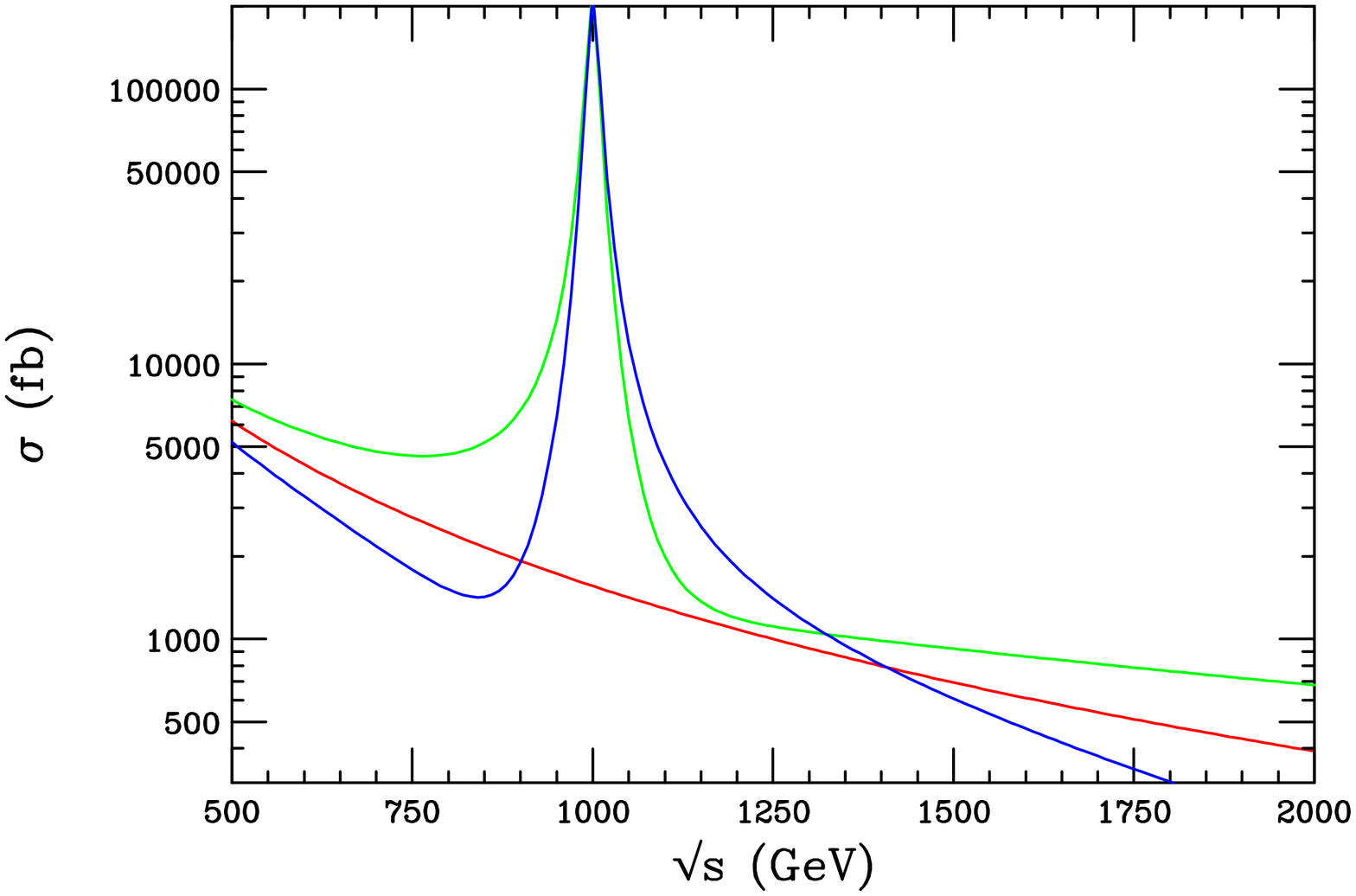}}
\vspace*{0.1cm}
\centerline{
\includegraphics[width=7.5cm,angle=90]{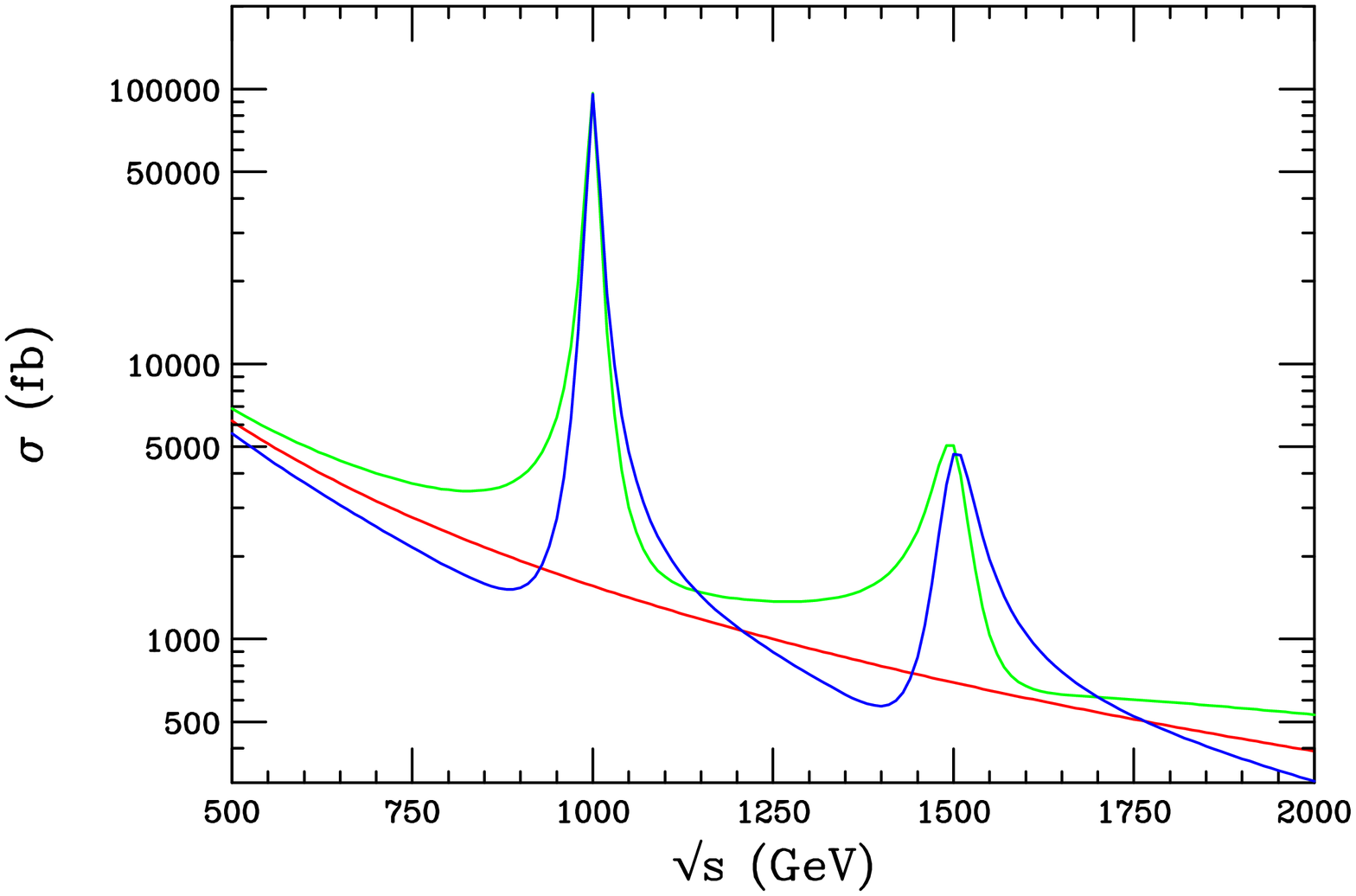}}
\vspace*{0.1cm}
\caption{Cross section for $e^+e^- \to e^+e^-$ as a function of $\sqrt s$ for scenarios $(a)$ and $(c)$ in green and for scenario $(b)$ in blue. The SM prediction 
for comparison purposes is in red. In the top panel $M_{W^0}=M_B=1$ TeV whereas in the bottom panel $M_B=1$ TeV and $M_{W^0}=1.5$ TeV. A cut on $z$ has been applied, 
$z\leq 0.8$, to remove the large contribution from the forward photon pole.}
\label{fig2}
\end{figure}

\begin{figure}[htbp]
\centerline{
\includegraphics[width=7.5cm,angle=90]{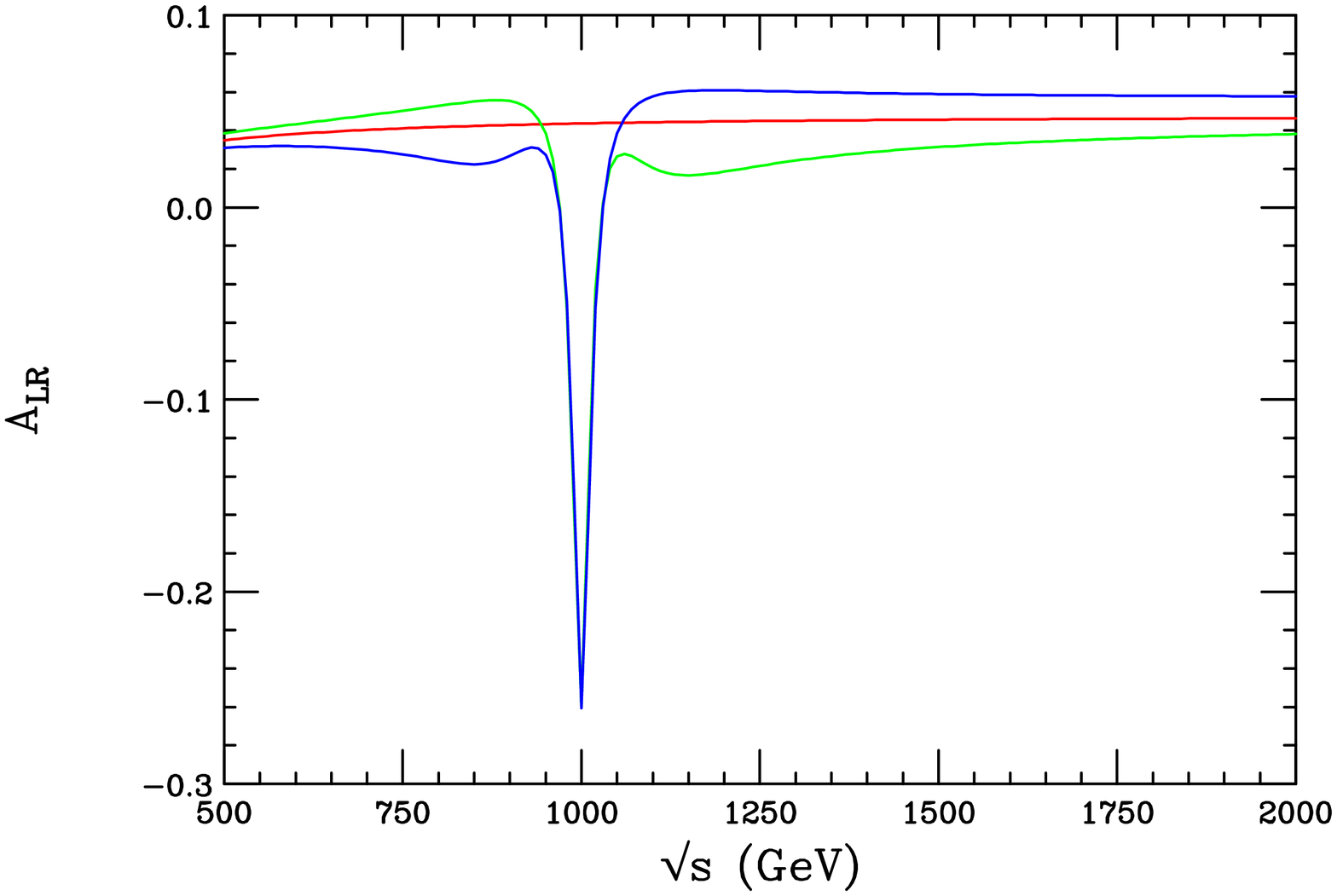}}
\vspace*{0.1cm}
\centerline{
\includegraphics[width=7.5cm,angle=90]{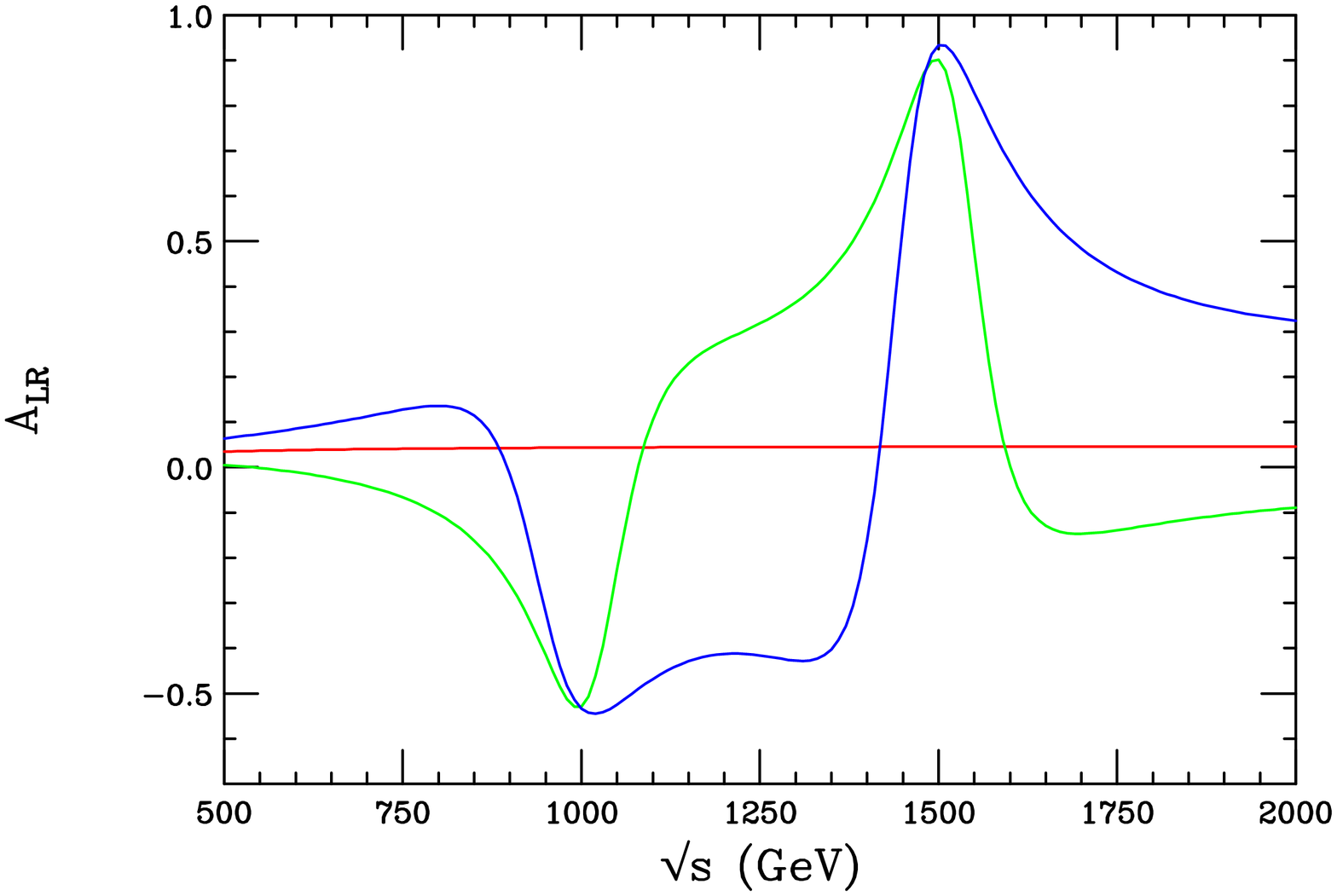}}
\vspace*{0.1cm}
\caption{Same as the previous figure but now for the angular averaged values of $A_{LR}^e$.}
\label{fig2b}
\end{figure}

At $e^+e^-$ colliders the situation is far simpler and we are directly led to consider Bhabha scattering, $e^+e^- \to e^+e^-$, which has {\it identical} 
initial and final state fermions so that we can no longer play the coupling sign trick. This process will depend upon coupling combinations like $B_{ij}$ and $C_{ij}$ 
above but with $f=e$. This means that a change in the sign of the 
electron's couplings to both $W^0$ and $B$ will leave the differential cross section and polarization asymmetries {\it invariant}. Explicitly we obtain
\begin{equation}
{{d\sigma}\over {dz}}={1\over {16\pi s}}\sum_{i,j}\Big[(B_{ij}+C_{ij})(P^{ss}_{ij}+2P^{st}_{ij}+P^{tt}_{ij}){{u^2}\over {s^2}}+(B_{ij}-C_{ij})(P^{ss}_{ij}{{t^2}
\over {s^2}} +P^{tt}_{ij})\Big]\,,
\end{equation}
where $t,u=-s(1\mp z)/2$ and, generalizing the above relation, we now write   
\begin{equation}
P^{qr}_{ij}=\lambda_{ij}s^2{{(q-M_i^2)(r-M_j^2)+\Gamma_i\Gamma_jM_iM_j}\over {[(q-M_i^2)^2+\Gamma_i^2M_i^2][(r-M_j^2)^2+\Gamma_j^2M_j^2]}}\,.
\end{equation}
From these expressions it is clear that for Bhabha scattering, cases $(a)$ and $(c)$ will yield identical cross section and asymmetry 
results while now case $(b)$, the GOW scenario, will be 
distinct. This is shown explicitly in Figs.~\ref{fig2} and ~\ref{fig2b} for the same parameter choices as employed above in Figs.~\ref{fig1} and ~\ref{fig1a}. 
Here we see that the previously obtained ambiguities have been removed and that the LW gauge fields can be uniquely identified as desired.

A weakness in the analysis above is that we may not have immediate access to a $e^+e^-$ collider with energies above 1 TeV so that it may be impossible to directly 
access the gauge boson excitation curves in Bhabha scattering, as we have done above, for some time. This depends upon, \eg, the potential relative schedules of the 
ILC and CLIC as well as many other known and unknown unknowns. However, it is clear that at the first stage of the  
ILC, we will likely be limited to values of $\sqrt s \leq 500$ GeV so that the properties of these new gauge bosons can only be indirectly studied in Bhabha scattering. 
Obviously this is a more difficult situation than in the case where the resonance region(s) of the new gauge bosons can be directly accessed. What can we learn at these 
lower energies below the resonances? Here the capability of the ILC to make precision measurements becomes of great importance. In the analysis below 
we will assume that the LHC has already determined the masses of the new gauge states and has made a relatively detailed study of their couplings 
to the SM fermions{\cite {tasi}}, determining that they are indeed SM-like.

\begin{figure}[htbp]
\centerline{
\includegraphics[width=7.5cm,angle=90]{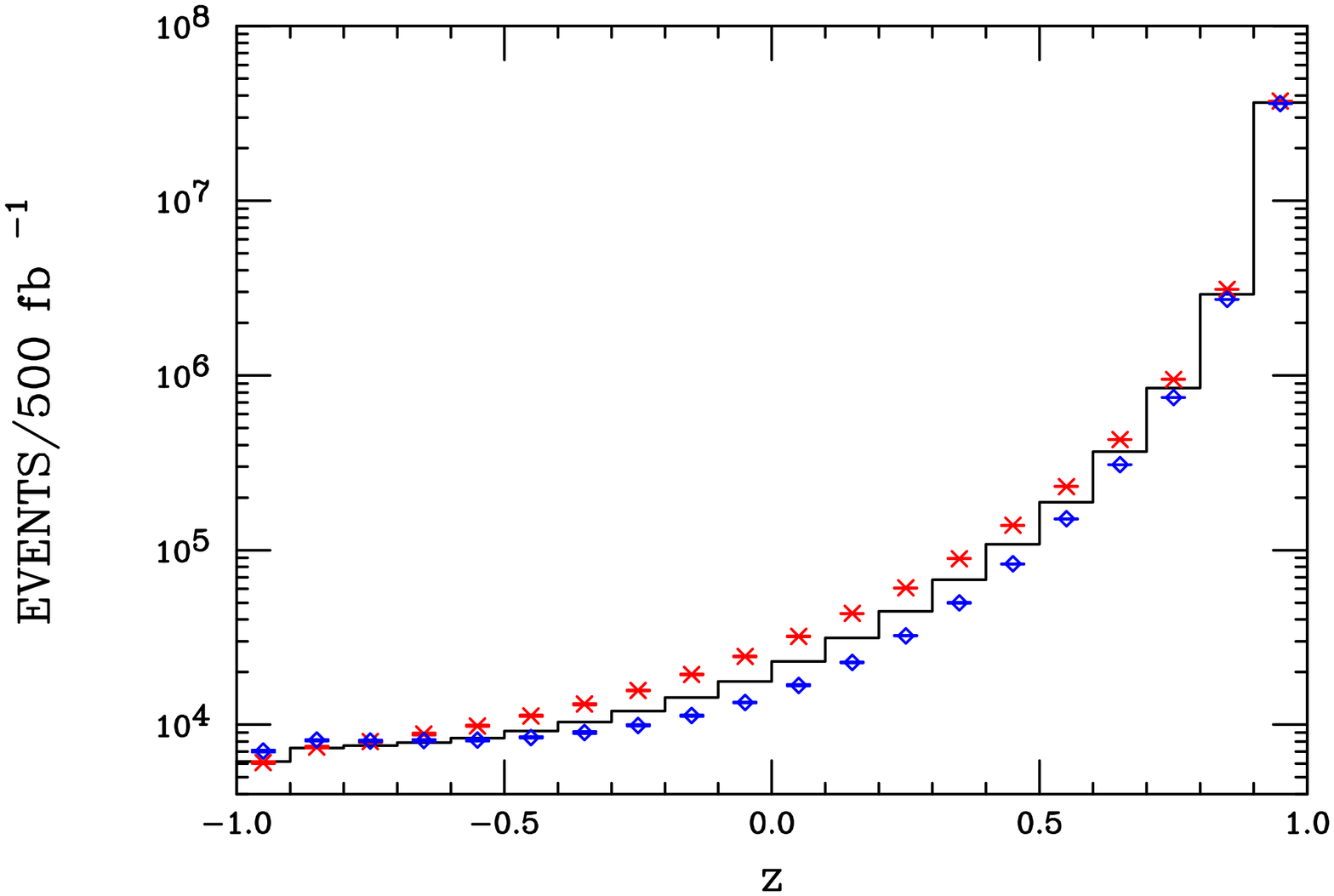}}
\vspace*{0.1cm}
\centerline{
\includegraphics[width=7.5cm,angle=90]{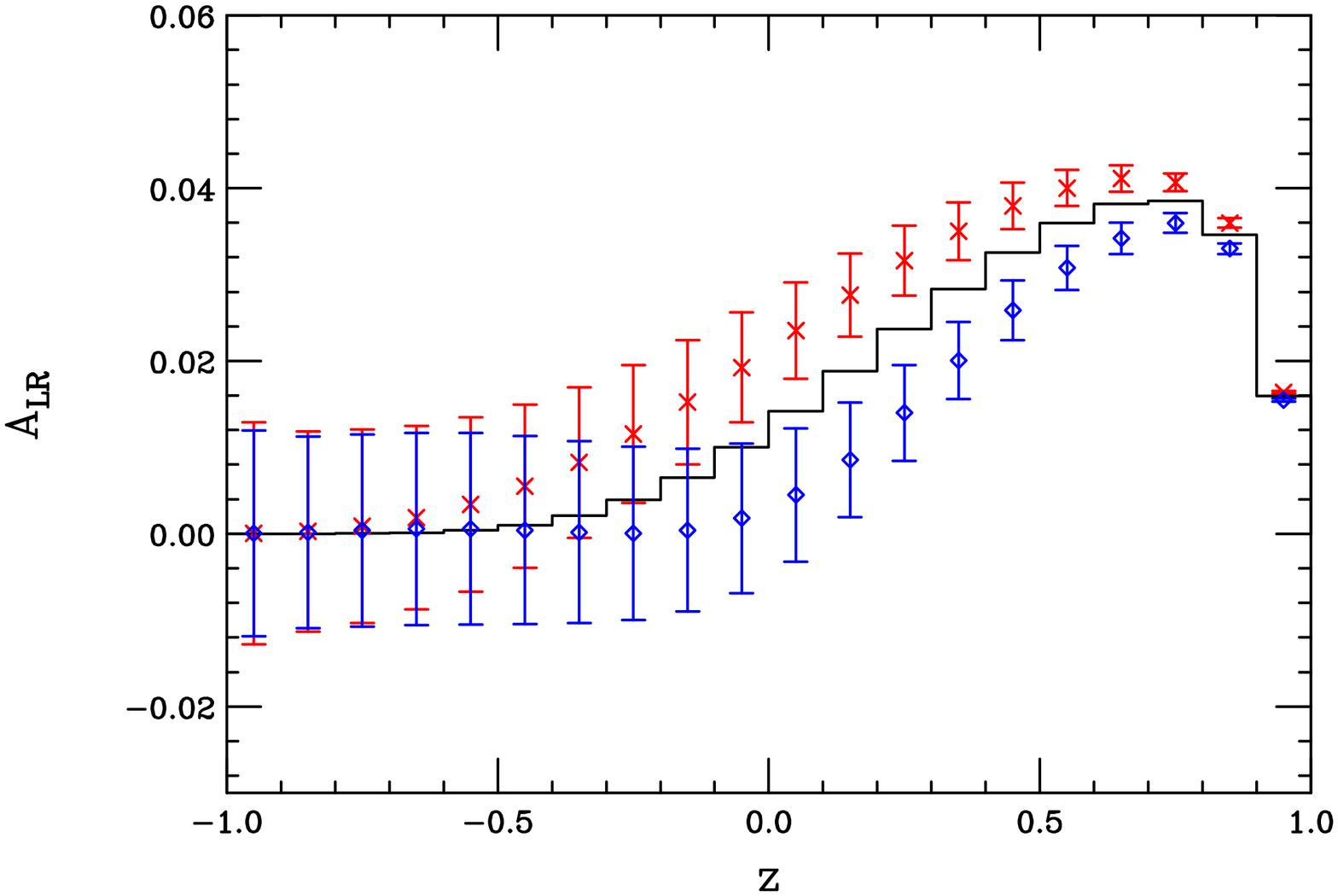}}
\vspace*{0.1cm}
\caption{Cross section and polarization asymmetry for $e^+e^- \to e^+e^-$ as a function of $z$ at $\sqrt s=500$ GeV for the various scenarios discussed in the text. 
The statistical errors in the measurements are shown. 
The black histogram is the SM result whereas the red data points are for scenarios $(a)$ or $(c)$; the blue ones are for the GOW model. Here, $M_{W^0,B}=1$ TeV has 
been assumed as well as an integrated luminosity of 500 $fb^{-1}$. ISR has been included with a cut on the $e^+e^-$ invariant mass $>400$ GeV;  
beamstrahlung effects have been ignored for simplicity.}
\label{fig3}
\end{figure}

\begin{figure}[htbp]
\centerline{
\includegraphics[width=7.5cm,angle=90]{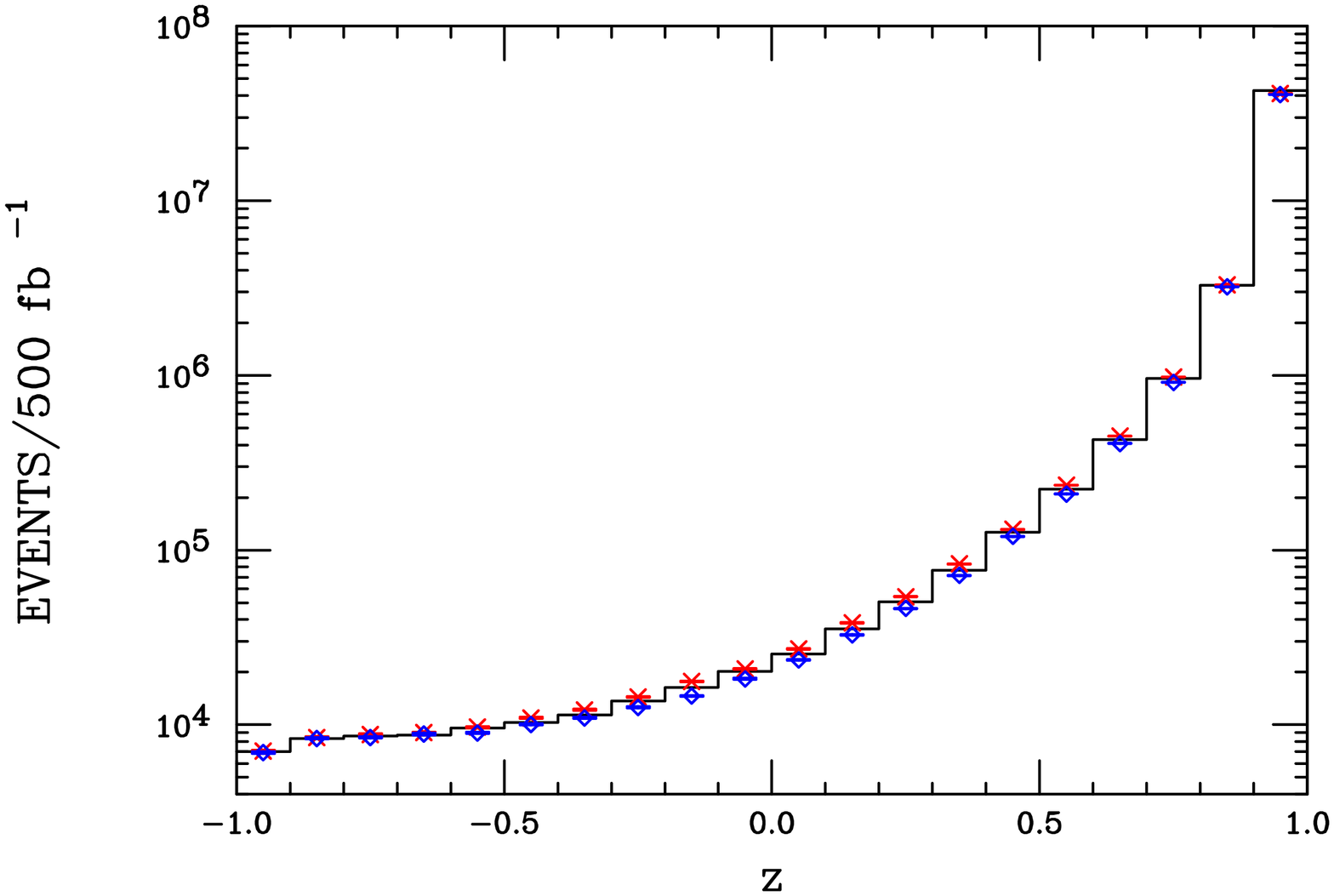}}
\vspace*{0.1cm}
\centerline{
\includegraphics[width=7.5cm,angle=90]{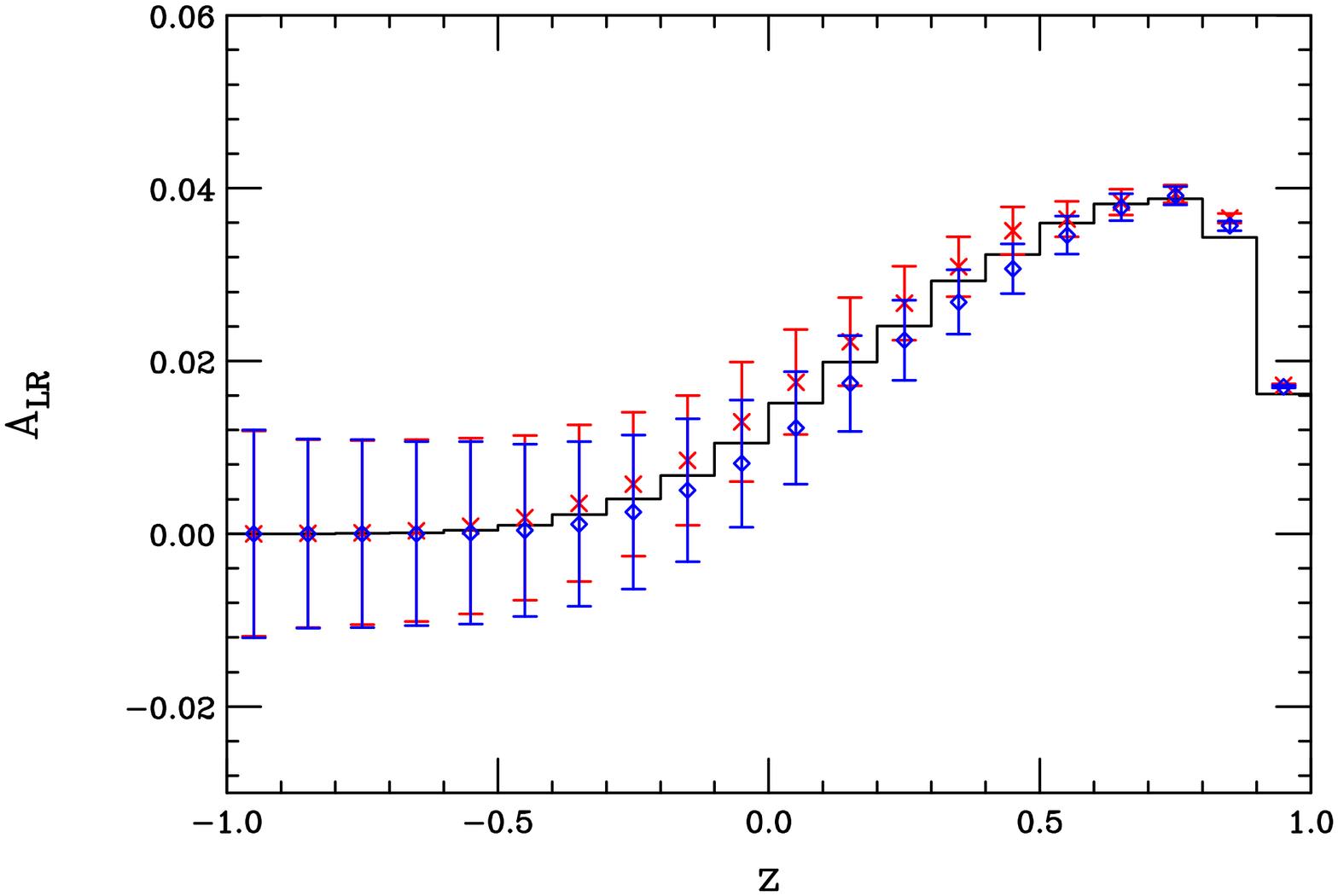}}
\vspace*{0.1cm}
\caption{Same as the previous figure but here $M_{W^0,B}=2$ TeV has been assumed.}
\label{fig4}
\end{figure}

Fig.~\ref{fig3} shows the results of this 500 GeV ILC analysis below the resonance region where it has been assumed that $M_{W^0,B}=1$ TeV. Away from the 
forward and backward directions it is quite clear that identical constructive 
interference occurs for scenarios $(a)$ and $(c)$ while destructive interference occurs for the GOW case $(b)$. At this level of statistics, these 
two possibilities are now very easily distinguished 
in both the differential cross section as well as in $A_{LR}^e(z)$. Of course, as the masses of the two fields $W_{LW}^0$ and $B_{LW}$ are increased this distinguishing  
power goes down quite rapidly as can be seen in Fig.~\ref{fig4} where it has now been assumed that $M_{W^0,B}=2$ TeV. Here we see that the two predictions are somewhat 
closer but are still separable given the large statistics. Certainly once we reach  $W^0,B$ masses of order $\simeq 3$ TeV and above, at these assumed integrated 
luminosities, this separation is no longer possible and a higher energy $e^+e^-$ collider will be required. In fact, we find that the overall separation reach scales 
roughly as $M \simeq 5\sqrt s$ for analyses performed below the LW resonance region. 


\section{Discussion and Conclusions}

In this paper we have demonstrated that the neutral, negative-normed gauge boson states predicted by the GOW model can be uniquely identified as such at future $e^+e^-$ 
colliders through the Bhabha scattering channel over a reasonably wide kinematic range. This overcomes the identification problem encountered for LW-type gauge bosons 
encountered by using data from the LHC alone. For $e^+e^-$ colliders with direct access to the (multi-)TeV scale associated with the resonance region(s) of these 
states, this identification is rather straightforward by using both cross section and polarization symmetry information that can be easily obtained. However, we also 
demonstrated that even at energies a factor of a few below such resonance masses, precision measurements of these same observables at $e^+e^-$ colliders can be used 
to uniquely identify the LW nature of new states provided these gauge boson masses are already known from LHC data and provided sufficient integrated luminosity is 
available. We this conclude that with data from $e^+e^-$ colliders the ambiguity issues associated with the production of LW gauge bosons can be easily resolved.

%
\def\MPL #1 #2 #3 {Mod. Phys. Lett. {\bf#1},\ #2 (#3)}
\def\NPB #1 #2 #3 {Nucl. Phys. {\bf#1},\ #2 (#3)}
\def\PLB #1 #2 #3 {Phys. Lett. {\bf#1},\ #2 (#3)}
\def\PR #1 #2 #3 {Phys. Rep. {\bf#1},\ #2 (#3)}
\def\PRD #1 #2 #3 {Phys. Rev. {\bf#1},\ #2 (#3)}
\def\PRL #1 #2 #3 {Phys. Rev. Lett. {\bf#1},\ #2 (#3)}
\def\RMP #1 #2 #3 {Rev. Mod. Phys. {\bf#1},\ #2 (#3)}
\def\NIM #1 #2 #3 {Nuc. Inst. Meth. {\bf#1},\ #2 (#3)}
\def\ZPC #1 #2 #3 {Z. Phys. {\bf#1},\ #2 (#3)}
\def\EJPC #1 #2 #3 {E. Phys. J. {\bf#1},\ #2 (#3)}
\def\IJMP #1 #2 #3 {Int. J. Mod. Phys. {\bf#1},\ #2 (#3)}
\def\JHEP #1 #2 #3 {J. High En. Phys. {\bf#1},\ #2 (#3)}

\end{document}